\documentclass[a4paper,fleqn,usenatbib]{mnras}
\usepackage{threeparttable}
\usepackage{tabularx}

\usepackage{newtxtext,newtxmath}
\usepackage{mathptmx}
\usepackage{txfonts}

\usepackage[T1]{fontenc}
\usepackage{aecompl}

\usepackage{graphicx}	% Including figure files
\usepackage{amsmath}	% Advanced maths commands
\usepackage{amssymb}	% Extra maths symbols

\title[Line Distributions of Seyfert 2]{Emission Line Luminosity Distributions of Seyfert 2 Galaxies}

\author[Chen et al.]{Yen-Chen Chen,$^{1,2,3}$\thanks{E-mail: m1029006@gm.astro.ncu.edu.tw}
 Chorng-Yuan Hwang$^{1}$
\\
$^{1}$Graduate Institute of Astronomy, National Central University, Chung-Li 32054, Taiwan\\
$^{2}$ICRA and Dipartimento di Fisica, Sapienza Università di Roma, Piazzale Aldo Moro 5, I-00185 Rome, Italy\\
$^{3}$ICRANet, Piazza della Repubblica 10, I-65122 Pescara, Italy}

\date{Accepted XXX. Received YYY; in original form ZZZ}

\pubyear{2018}

\usepackage{etoolbox}
\makeatletter
\patchcmd\@combinedblfloats{\box\@outputbox}{\unvbox\@outputbox}{}{%
   \errmessage{\noexpand\@combinedblfloats could not be patched}%
}%
 \makeatother

\begin{document}
\label{firstpage}
\pagerange{\pageref{firstpage}--\pageref{lastpage}}
\maketitle

% Abstract of the paper
\begin{abstract}
We probed the relation between line activities of Seyfert 2 galaxies and their host galaxies. We selected Seyfert 2 galaxies from the Sloan Digital Sky Survey Data Release 10 with redshifts less 0.2. We used the luminosity of the emission lines as indicators of AGN power. 
%We found that there is a correlation between the bulge magnitudes of the host galaxies of the Seyfert 2 and the emission line  luminosities of the Seyfert 2. Besides, 
We found that the Seyfert 2 galaxies seem to have two populations in the emission line luminosity distributions. We considered the $L_\mathrm{[OIII]}/L_{\textnormal{bulge}}$ ratio as an accretion rate indicator and found that the two Seyfert 2 distributions seem to have different accretion rates. We found that these two Seyfert 2 populations, although classified by their emission line distributions, turned out to have different morphology distributions. 
%We also found that a small fraction of the Seyfert 2 galaxies have relative young stellar populations and their host galaxies are dominated by late-type galaxies, which is distinct from the host galaxies of most Seyfert 2 galaxies. 
These results indicate that these different populations of the Seyfert 2 galaxies might be significantly different in their physical conditions.
\end{abstract}

%We found that the Seyfert 2 galaxies with $\log L_{\mathrm{[OIII]}}~\mathrm{[ergs/s]}~<~40.125$ show a correlation between their [OIII] and continuum emission whereas there is no such a correlation for the Seyfert 2 galaxies with $\log L_{\mathrm{[OIII]}}~\mathrm{[ergs/s]}~>~40.125$.

% Select between one and six entries from the list of approved keywords.
% Don't make up new ones.

%%%%%%%%%%%%%%%%%%%%%%%%%%%%%%%%%%%%%%%%%%%%%%%%%%

%%%%%%%%%%%%%%%%% BODY OF PAPER %%%%%%%%%%%%%%%%%%

\section{Introduction}

According to the traditional unification model of Seyferts \citep{Antonucci93, Urry95}, Seyfert 1 and Seyfert 2 galaxies are intrinsically the same. The difference between Seyfert 1 and Seyfert 2 galaxies is only caused by the viewing orientation. Seyfert 1 galaxies are viewed face-on relative to their tori and accretion disks whereas Seyfert 2 galaxies are viewed edge-on. The polarized optical spectra of some Seyfert 2 galaxies showed broad emission lines and suggested that these Seyfert 2 galaxies have hidden broad line regions \citep{Antonucci85,Antonucci93}; these results strongly support the traditional unification model. However,  some other Seyfert 2 galaxies did not show broad emission lines in the polarized optical spectroscopic observations \citep{Tran03, Nicastro03}. The Seyfert 2 galaxies with broad emission lines in the polarized spectrum are considered as hidden broad line region (HBLR) Seyfert 2 galaxies, whereas the Seyfert 2 galaxies without broad emission lines in the polarized spectrum are called as non-HBLR Seyfert 2s \citep{Tran95,Heisler97,Tran01}. The different characteristic in the optical spectra of the Seyfert 2 galaxies suggests that there might be different origins for the Seyfert 2 activities. 

It was unclear what physical mechanisms generate these different Seyfert 2 phenomena. The optical spectral features of Seyfert 2 galaxies contain many different narrow emission lines with both high- and low-ionization. Most of these emission lines are believed to be caused by photoionized gas surrounding the super massive black hole (SMBH) in the galaxy center \citep{Rees84,Veilleux87}. However, there are another possible explanations to account for the phenomena of Seyfert 2 galaxies; for example, \citet{Terlevich85} examined the effect of a number of young hot stars in the nucleus of Seyfert galaxies and suggested that the activity power of Seyfert 2 galaxies is associated with extreme star-formation rate. Besides, \citet{Kraemer98} suggested that the mechanism for Seyfert 2 galaxies can be explained by dominant photoionization with an additional heating of collision by shock after comparing ultraviolet and optical spectra for different regions within 100 pc of the nucleus of NGC 1068 with the results of emission line ratios from photoionization models. \citet{Watabe08} found that the correlation between nuclear starburst luminosity and AGN luminosity is tighter than that between circumnuclear starburst luminosity and AGN luminosity, indicating that central starburst of AGN is related to AGN mass accretion.

The physical origin for the presence of both high- and low- ionization emission lines in the optical spectra of Seyfert 2 galaxies is still difficult to reproduce with a single model. The observed emission line ratios of Seyfert 2 galaxies could be modeled by fast-velocity radiative shocks, while the low-ionization LINERs might be produced by shocks with low velocities \citep{Dopita95}. On the other hand, photoionization models taking into account the effect of dust succeeded to reproduce observed ratios of emission-lines for Seyfert 2 galaxies with high ionization parameters \citep{Groves04}. Investigating the emission lines in the spectra of Seyfert 2 galaxies will help us to understand the physical process of the Seyfert 2 galaxies.

In this paper, we investigate relationship between the emission line activity of the Seyfert 2 galaxies and their host galaxies to probe the origin of Seyfert 2 phenomena. We have selected more than 30000 Seyfert 2 galaxies from the Sloan Digital Sky Survey (SDSS) for statistical analysis. In Section 2, we describe the selection of our Seyfert 2 samples. In Section 3, we present the relationship between bulge magnitudes and the luminosities of different emission lines for our Seyfert 2 samples. In Section 4, we present the relationship between the accretion rate indicator $L_\mathrm{[OIII]}/L_{\textnormal{bulge}}$ and the stellar age indicator $D_n$(4000) for our Seyfert 2 galaxies. Finally, we discuss and summary our results in Section 5 \& 6. In this paper, we used $H_{0}$=70~km~s$^{-1}$~Mpc$^{-1}$, $\Omega_{m}=0.3$, $\Lambda_{0}=0.7$, $q_{0}=-0.55$, $k=0.00$.

\section{Sample Selection}

We selected the Seyfert 2 sample from the SDSS DR10 \citep{Ahn14}.
We first considered the SDSS sources that have both photometry and spectrum data and were classified as ``GALAXY''  in the SDSS photometry. We only selected sources with redshift less than 0.2 in order to cover some important emission lines within the spectral range. There are 738180 SDSS galaxies in the redshift range. We then selected Seyfert 2 galaxies using the following criteria:

\begin{itemize}
\item Criteria of \citet{Kewley06}
\begin{itemize}
\item $\mathrm{\log([OIII]/H\beta)~>~1.3~+~0.72/ [\log([SII]/H\alpha)-0.32]}$
\item $\mathrm{\log([OIII]/H\beta)~>~0.76~+~1.89\log([SII]/H\alpha)}$
\end{itemize}
\item Full width at half maximum (FWHM) of the Balmer lines $< 1000~\mathrm{km~s^{-1}}$
\end{itemize}
\begin{itemize}
\item Signal-to-noise ratio~$>$~3 of [OIII] and H$\alpha$ emission lines 
\end{itemize}

The distribution of [SII]/H$\alpha$ showed a bimodal distribution separating LINER from Seyfert 2 galaxies according to \citet{Kewley06}. Therefore, we used these two criteria listed above to separate Seyfert 2 galaxies from star-forming galaxies and LINERs. Finally, we have 30768 Seyfert 2 galaxies.

\section{Bulge Magnitudes and Emission Lines}
We investigated the relationship between bulge magnitudes and the luminosities of emission lines for our Seyfert 2 samples. Bulge magnitudes were derived by using the parameters derived from SDSS galaxy fitting. The galaxy was fitted by the
``{\tt{cmodel}}'' that is a combination of a de Vaucouleurs flux and an exponential flux.
The de Vaucouleurs profile was used to fit the bulge of the galaxy and the exponential profile was used to fit the disk of the galaxy.
We used the cmodel flux to represent the flux of the whole galaxy and the bulge flux were derived from the cmodel flux multiplied by {\it{FracDev}}, which is a coefficient to represent the fraction of the de Vaucouleurs term in the host galaxy in the SDSS fitting, to get the bulge magnitude of our galaxy samples. We considered the emission lines of [OIII]$\lambda5007$ and $\mathrm{H\alpha}$ to represent the AGN activities; the fluxes of the emission lines were obtained from table of the galSpecLine in the SDSS DR10 and derived by the methods of \citet{Tremonti04} and \citet{Brinchmann04}.

\begin{figure}
\begin{center}
\includegraphics[width=0.45\textwidth]{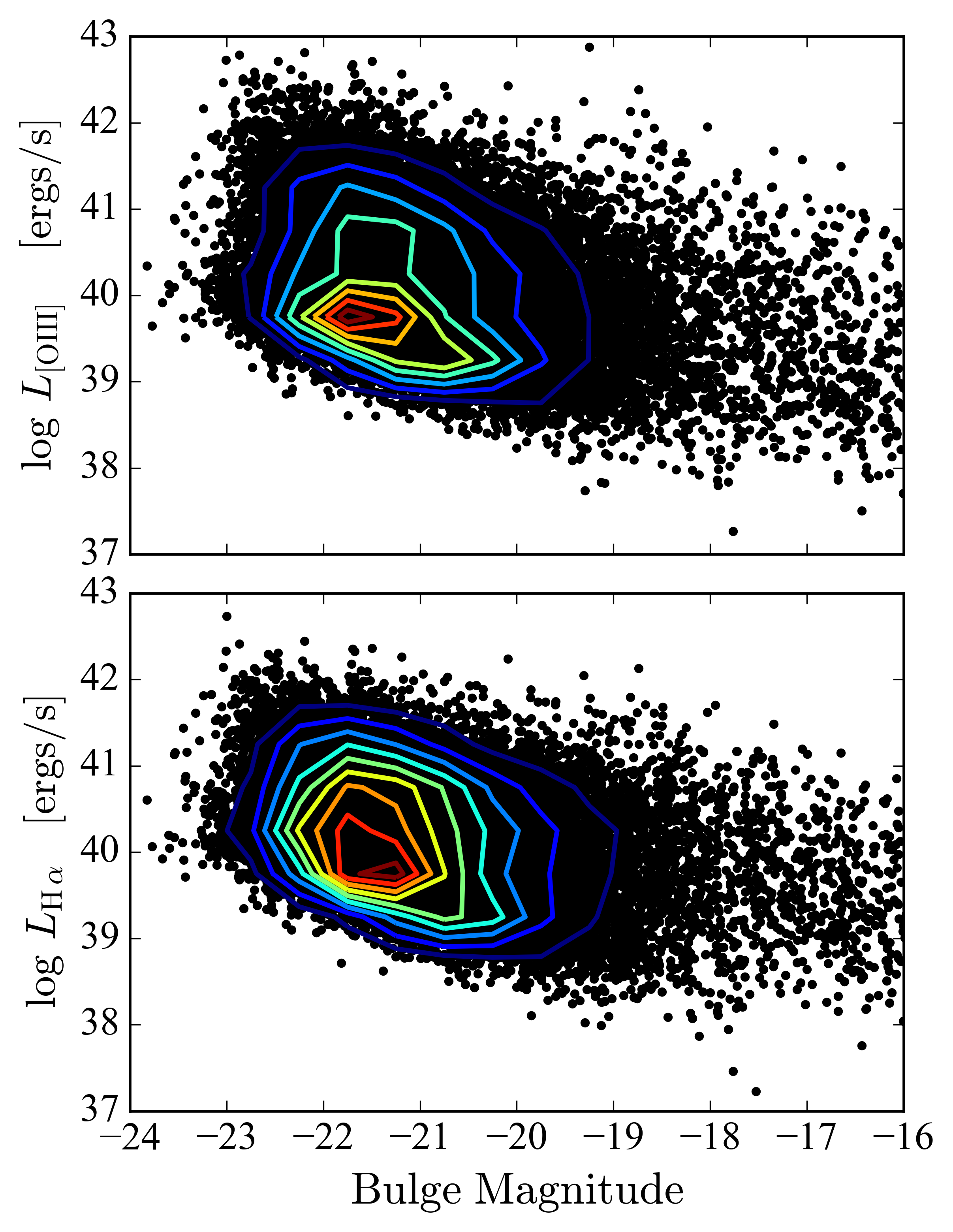}
\caption{ Bulge magnitudes $(M_\mathrm{bulge})$ versus emission line luminosities for the Seyfert 2 galaxies. Top: Bulge magnitude versus $L_\mathrm{[OIII]}$. Bottom: Bulge magnitude versus $L_\mathrm{H\alpha}$. Pink lines represent the best linear fitting in each panel. The dark red contour represents highest surface density of the sources in the distribution diagram; the red contour is 88\% of the dark red one, the orange 77\% of the dark red one, the green-yellow 66\% of the dark red one, the green 55\% of the dark red one, the aqua 44\% of the dark red one, the light blue 33\% of the dark red one, the blue 22\% of the dark red one, the navy 11\% of the dark red one}
\label{s1s2be}
\end{center}
\end{figure}

%\begin{table}
%\caption{Results of Pearson correlation test for Fig.~\ref{s1s2be}.}
%\label{table: pcor}
%\begin{tabular}{lccc}
%\hline
% & Correlation coefficient & P-value & Scatter\\
%\hline
%$\log L_\mathrm{[OIII]}$-{\bf{$M_\mathrm{bulge}$}} & -0.342  & 0 & 0.524\\
%$\log L_\mathrm{H\alpha}$-{\bf{$M_\mathrm{bulge}$}} & -0.365 & 0 & 0.408\\
%$\log L_\mathrm{[NII]}$-{\bf{$M_\mathrm{bulge}$}} & -0.59 & 2.22$\times10^{-245}$ & 0.21\\
%\hline
%\end{tabular}
%\end{table}

\begin{figure}
\begin{center}
\includegraphics[width=0.45\textwidth]{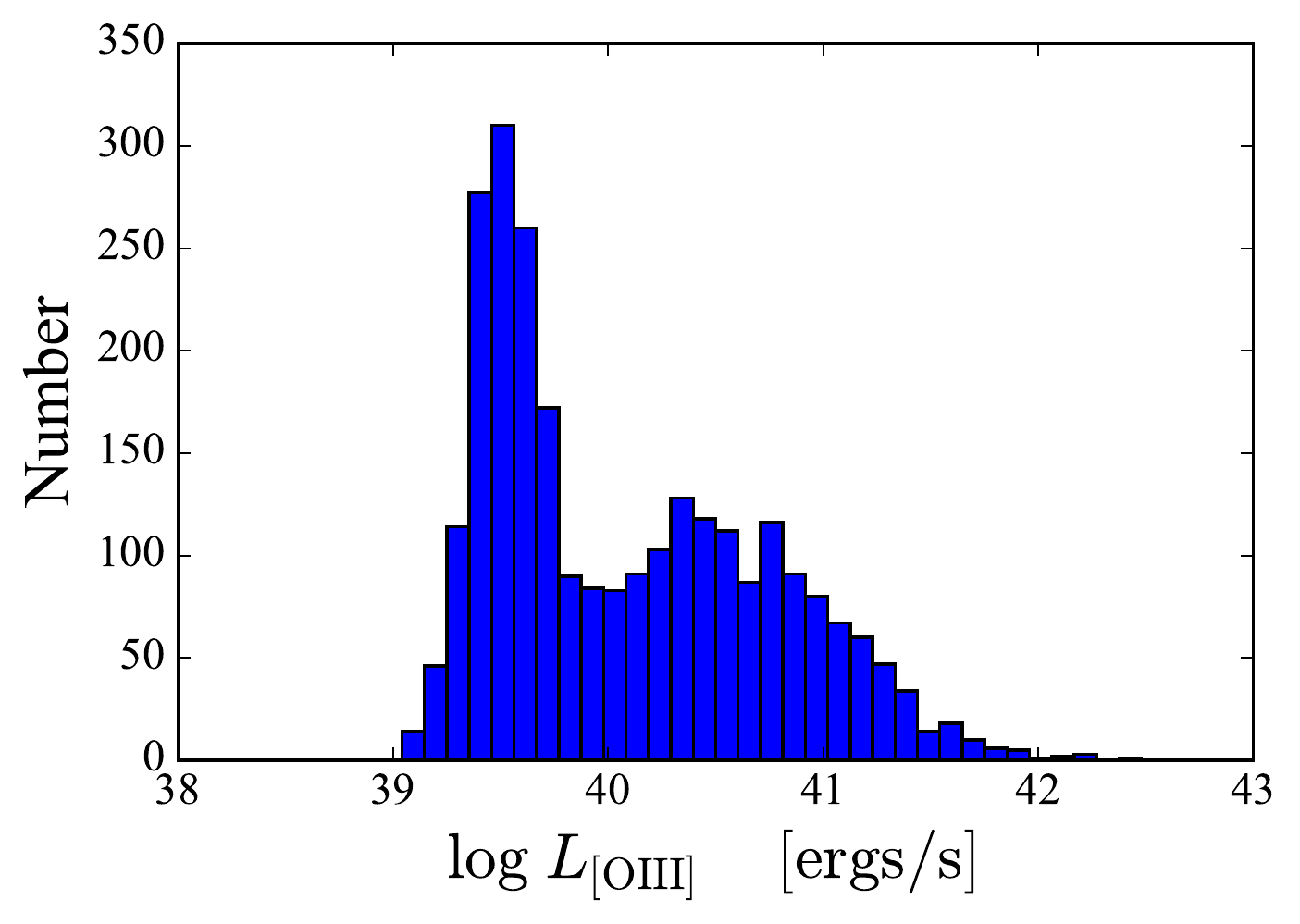}
\caption{$L_{\mathrm{[OIII]}}$ distribution for the sub-sample Seyfert 2 galaxies with $-22 <$ M$_{\mathrm{bulge}} < -21$.}
\label{sub_s2o3dV_bul}
\end{center}
\end{figure}

\begin{figure*}
\begin{center}
\includegraphics[width=0.8\textwidth]{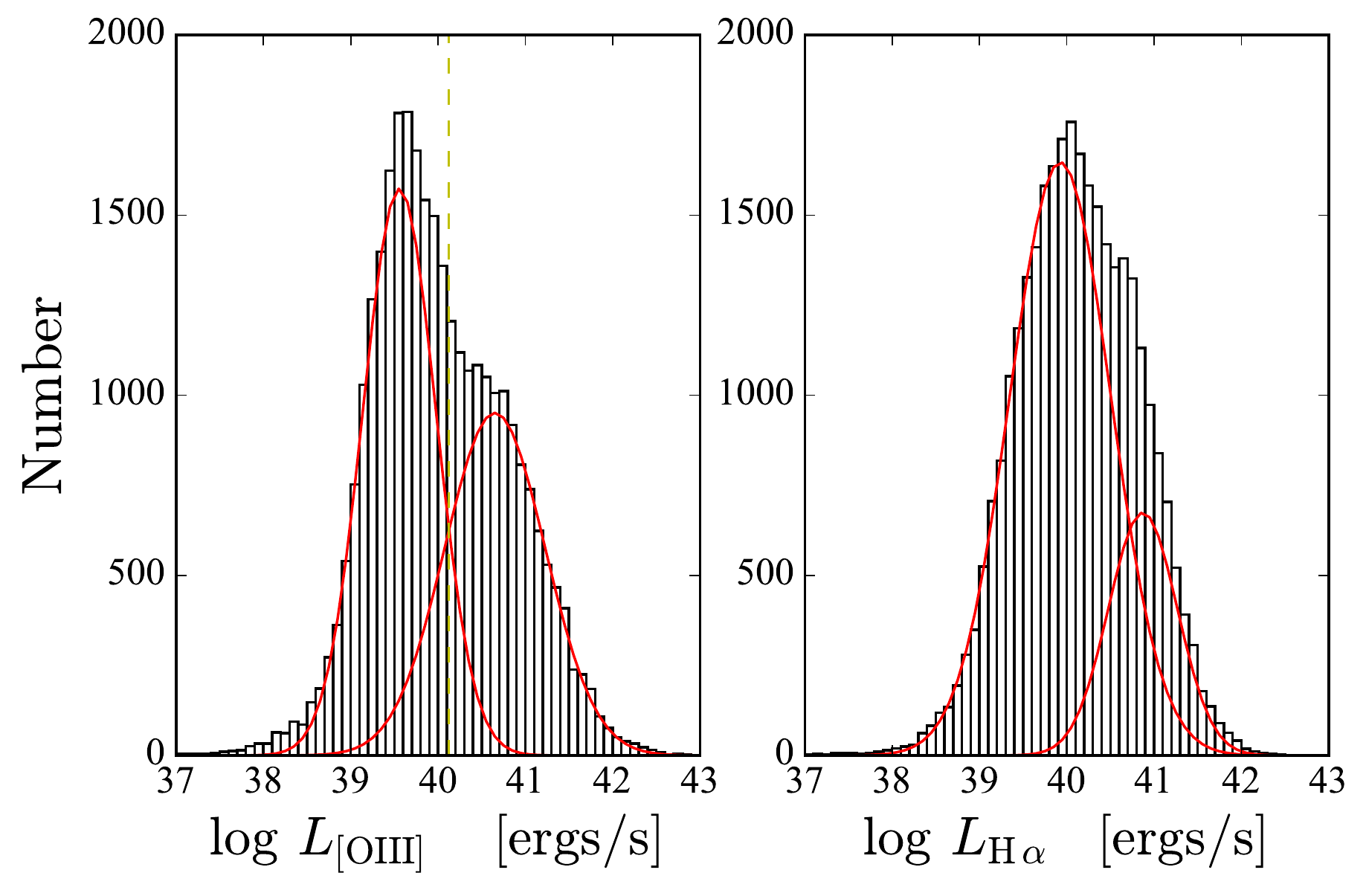}
\caption{Luminosity distributions of [OIII] and $\mathrm{H\alpha}$ emission lines for the Seyfert 2 galaxies. Left: $L_\mathrm{[OIII]}$ distributions. Right: $L_\mathrm{H\alpha}$ distributions. Green dashed line represents separation line of $\log L_{\mathrm{[OIII]}} \mathrm{[ergs/s]} = 40.125$. Red lines: fitting results of two-gaussian functions.}
\label{s2o3d}
\end{center}
\end{figure*}

Fig.~\ref{s1s2be} shows the bulge magnitudes {\bf $(M_\mathrm{bulge})$} versus the luminosities of emission lines for our Seyfert 2 galaxies. We found that brighter bulge magnitudes generally have stronger emission line luminosities; however, the surface density distribution of the sources seems to be extended along the line luminosity. To demonstrate the extended line distribution, we plotted the $L_\mathrm{[OIII]}$ for a sub-sample of Seyfert 2 galaxies with bulge magnitude between -22 and -21 in Fig.~\ref{sub_s2o3dV_bul}. The double-peaked and extended distribution in Fig.~\ref{sub_s2o3dV_bul} shows that there are more than one component in the $L_\mathrm{[OIII]}$ distribution of Seyfert 2 galaxies with similar bulge magnitudes. To further investigate the distributions of line luminosities in Seyfert 2 galaxies, we plotted the total distributions of the line luminosities of [OIII] and H$\alpha$ for all selected Seyfert 2 galaxies in Fig.~\ref{s2o3d}. We also found that the distributions are not gaussian and show an extending tail in the high luminosity end. There are some Seyfert 2 galaxies showing unusual stronger luminosity than most Seyfert 2 galaxies doing.

% for [OIII] and H$\alpha$. These results display that there exists a correlation between the luminosities of these  emission lines and the bulge magnitudes for our Seyfert 2 galaxies. We estimated the Pearson correlation coefficients between line luminosities and bulge magnitudes. The results are shown in Table~\ref{table: pcor}. The correlation coefficients indicate that the bulge magnitudes and the emission line luminosities are significantly correlated. This suggests that AGN activities are relating to the bulge magnitudes of their host galaxies. Besides, we noted that the correlation between $L_\mathrm{[OIII]}$ and bulge magnitude has larger scatter value than the correlation between $L_\mathrm{H\alpha}$ and bulge magnitude. However, we found that the Seyfert 2 galaxies seem to have an extended distribution along the y-axis for the [OIII] luminosity in Fig.~\ref{s1s2be}. In Fig.~\ref{s2o3d}, we plotted the distributions of the line luminosities of [OIII] and H$\alpha$ for the Seyfert 2 galaxies.} 

\begin{table}
\caption{Reduced $\chi^{2}$ test for Fig.~\ref{s2o3d}}
\label{table: rchi1}
\begin{tabular}{lcc}
\hline
 & $\log L_\mathrm{[OIII]}$ & $\log L_\mathrm{H\alpha}$ \\ 
\hline
One-gaussian function & 42.94 & 6.10 \\
Two-gaussian function & 1.63 & 1.30 \\
\hline
\end{tabular}
\end{table}

We fitted a single gaussian function to the distributions of the emission line luminosities. We did the chi-square ($\chi^{2}$) test to compare the distributions of the emission line luminosities with the fitting gaussian function. Table~\ref{table: rchi1}. shows the results of the reduced chi-square. The results of one-gaussian function show large reduced chi-square values, suggesting that the line distributions can not be well described by a single gaussian function. We also fitted a two-gaussian function to the distributions of the emission line luminosities. We found that the reduced chi-square values for the two-gaussian fits can improve significantly comparing with the results of the one-gaussian function. We found that the reduced chi-square values suggest that the two-gaussian function fitting for the [OIII] luminosity distributions is acceptable. Therefore, we assumed that there are two populations in the Seyfert 2 sample; and we used the two-gaussian function fitting to separate the Seyfert 2 into high and low luminosity populations and tried to investigate the possible difference of their physical properties.

\section{Accretion Rate Indicator and $D_\mathrm{n}$(4000) Index}
To investigate the origin of these two populations, we divided our Seyfert 2 galaxies into high and low luminosity samples using {\bf $\log L_\mathrm{[OIII]} \textnormal{[ergs/s]} = 40.125$} as the separation luminosity. Therefore, we have 18015 low luminosity Seyfert 2 galaxies and 12753 high luminosity Seyfert 2 galaxies.

The separation line of $\log L_{\mathrm{[OIII]}} \mathrm{[ergs/s]}$ at 40.125 is about 1.34$\sigma$ away from the peak value of the low luminosity gaussian component, corresponding to the cut line of 9\%. In other word, the sources from the low luminosity gaussian with $\log L_{\mathrm{[OIII]}} \mathrm{[ergs/s]}$ greater than 40.125 are expected to be less than 9\% (or less than 1781 sources). Therefore, more than  86.3\% ((12753-1781)/12753=0.8603) of the sources with $\log L_{\mathrm{[OIII]}} \mathrm{[ergs/s]} >  40.125$ should be real sources from a different population. On the other hand, the mean value and the standard deviation of $\log L_{\mathrm{[OIII]}}$ for the high-luminosity component are 40.65 and 0.57. The separation at $\log L_{\mathrm{[OIII]}} \mathrm{[ergs/s]} = 40.125$ is about 0.9 $\sigma$ away from the mean of the high-luminosity component, suggesting that about 19\% or 2991 sources within the the high luminosity component might be classified as low luminosity sources. Below the separation luminosity, there are 18015 sources. Therefore, about 83.4\% ((18015-2991)/18015) sources should be from the real low luminosity components.

%18015 + 12753 = 30768

%1.34 sigma ~ 81.976%
%(100 - 82 ) / 2 = ~9%
% 18015 correspond to 91% low luminous gaussian
% expect gaussian = 18015/91*100=19797
% expect # = 19797*0.09=1781

%0.9 sigma ~ 63.188%
%(100 - 82 ) / 2 = ~19%
% 12753 correspond to 81% low luminous gaussian
% expect gaussian = 12753/81*100=15744
% expect # = 15744*0.19=2991

\begin{figure}
\begin{center}
\includegraphics[width=0.4\textwidth]{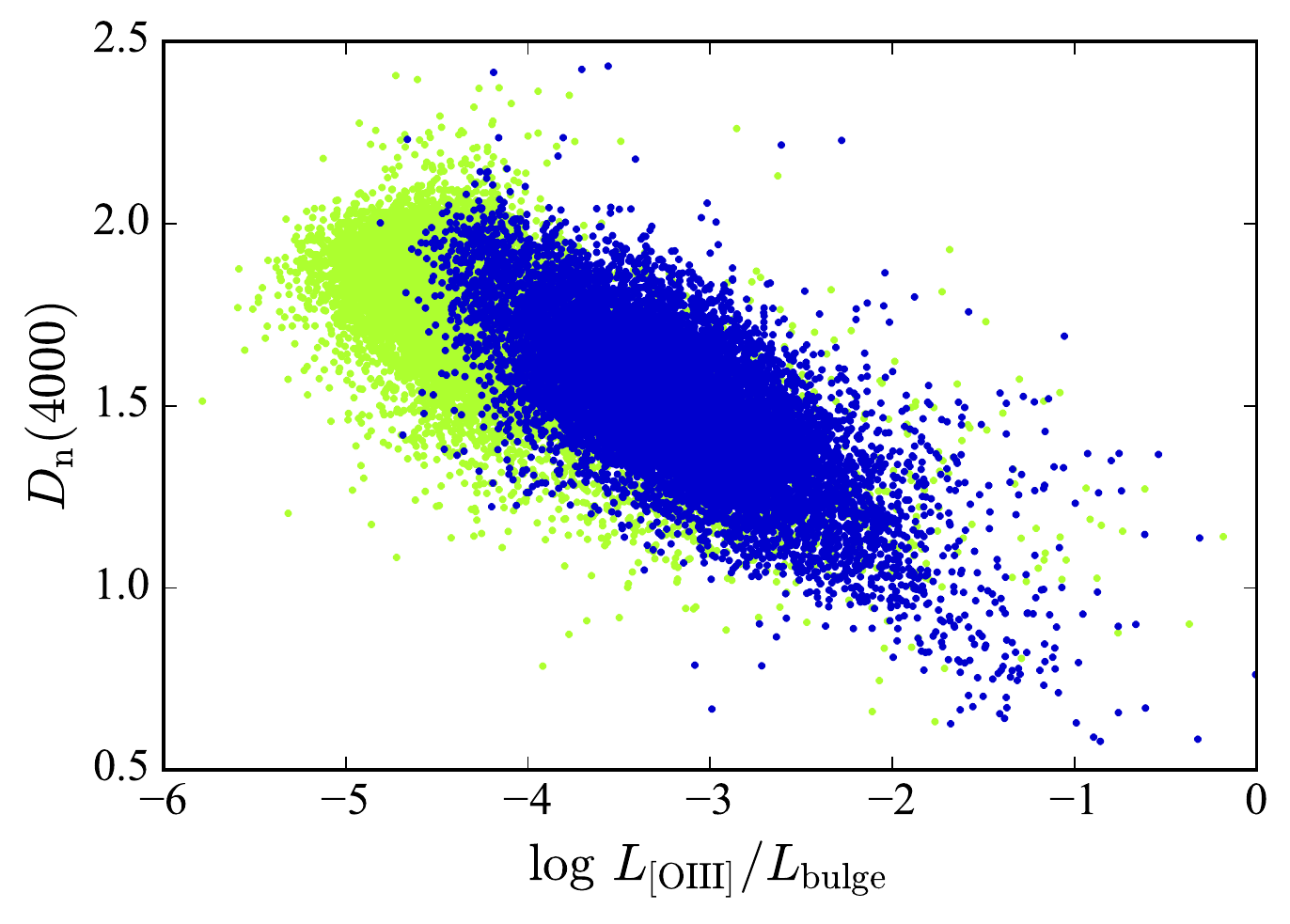}
\caption{$D_{\rm n}$(4000) versus ratios of $L_\mathrm{[OIII]}/L_{\textnormal{bulge}}$. Green dots represent the Seyfert 2 samples with $\log L_{\mathrm{[OIII]}} \mathrm{[ergs/s]} < 40.125$ and blue dots represent the Seyfert 2 samples with $\log L_{\mathrm{[OIII]}} \mathrm{[ergs/s]} > 40.125$.}
\label{s2d4000}
\end{center}
\end{figure}

The luminosity of AGN is proportional to its accretion rate and its black hole mass. The black hole mass of  an AGN is proportional to its bulge magnitude \citep{Kormendy95}, and $L_\mathrm{[OIII]}$ is related to the AGN power \citep{Heckman04}. Therefore, we use $L_\mathrm{[OIII]}/L_\mathrm{bulge}$ as the relative accretion rate of the Seyfert 2 galaxies. We also probed the stellar populations of our Seyfert 2 galaxies using the $D_n$(4000) strengths, which were defined by \citet{Bruzual83} and \citet{Balogh99} as:
\[D_n(4000)= \frac{\int_{4000}^{4100}f(\lambda)d\lambda}{\int_{3850}^{3950}f(\lambda)d\lambda}\]
The value of $D_n$(4000) can be considered as an age indicator of the stellar populations in the central regions of the host galaxies \citep{Kauffmann03a}. We obtained the $D_n$(4000) values of our Seyfert 2 sources from the SDSS databases \citep{Brinchmann04}. Fig.~\ref{s2d4000} shows the relation between $L_\mathrm{[OIII]}/L_{\textnormal{bulge}}$ and $D_n$(4000) in the Seyfert 2 galaxies. The Seyfert 2 galaxies with larger values of $L_\mathrm{[OIII]}/L_{\textnormal{bulge}}$ tend to have smaller values of $D_n$(4000). Besides, the Seyfert 2 galaxies with high $L_\mathrm{[OIII]}$ seem to concentrate around low values of $D_n$(4000). The low values of $D_n$(4000) might be caused by young stellar populations or affected by AGN continuum \citep[e.g.,][]{Yu13}.

\section{Discussion}

\begin{figure*}
\begin{center}
\includegraphics[width=0.7\textwidth]{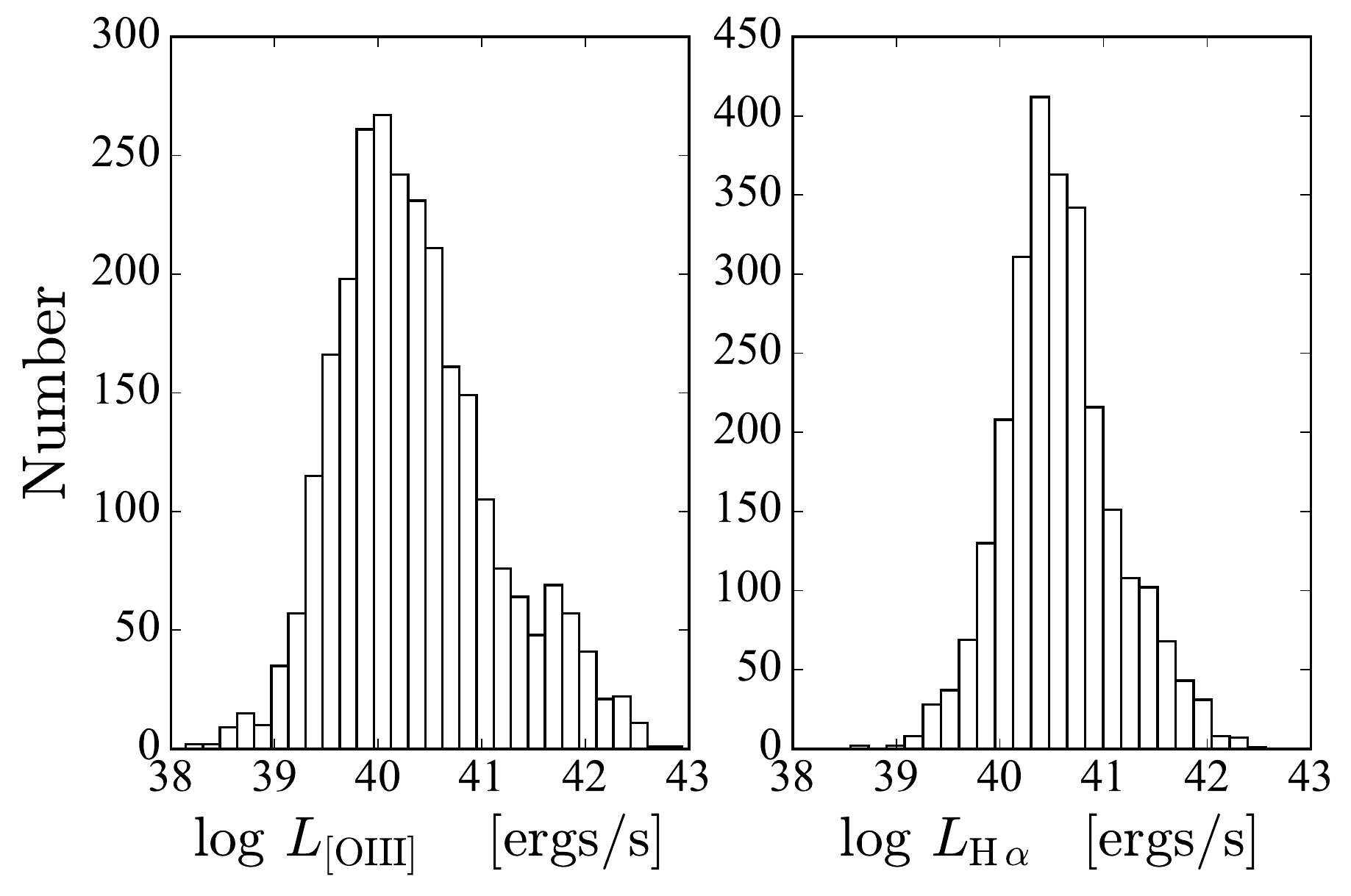}
\caption{Luminosity distributions of [OIII] and $\mathrm{H\alpha}$ emission lines for the Seyfert 2 galaxies identified by the V\'{e}ron catalog. Left: $L_\mathrm{[OIII]}$ distributions. Right: $L_\mathrm{H\alpha}$ distributions.}
\label{s2o3dV}
\end{center}
\end{figure*}

Fig.~\ref{s2d4000} demonstrates that the Seyfert 2 galaxies with larger accretion rates have smaller values of $D_n$(4000). This result might suggest that the Seyfert 2 galaxies with younger stellar populations in the host galaxies have larger accretion rates \citep[e.g.,][]{Heckman04, Wang15}. \citet{Kewley06} showed a similar trend for $D_n$(4000) versus $L_\mathrm{[OIII]}/\sigma^4$ in the Seyfert 2 galaxy samples selected from SDSS-DR4 with redshifts from 0.04 to 0.1. We noted that the Seyfert 2 galaxies of high $L_{\mathrm{[OIII]}}$ have smaller $D_n$(4000) values than the Seyfert 2 sources of \citet{Kewley06}. These high $L_{\mathrm{[OIII]}}$ Seyfert 2  sources have typical $D_n$(4000) values around $\approx$ 1, which is even lower than the lowest $D_n$(4000) values of the Seyfert 2 galaxies in \citet{Kewley06}. These results suggest that the low values of $D_n$(4000) in the high $L_{\mathrm{[OIII]}}$ sources might be also affected by some other different mechanisms, such as the AGN continuum \citep[e.g.,][]{Yu13} and can not be simply explained with the young stellar populations in the host galaxies. Furthermore, we notice that both the low and high luminous Seyfert 2 galaxies possess sources with $D_n$(4000) from 1.5 to 2.0, but these sources have different ratios of $L_\mathrm{[OIII]}/L_{\textnormal{bulge}}$ in these two populations, suggesting that even these two populations have similar stellar populations in their host galaxies, they might still have different accretion powers.

To check whether our results are affected by our selection criteria, we also selected the Seyfert 2 galaxies identified by the V\'{e}ron Catalog (13th) from the Sloan Digital Sky Survey (SDSS) with redshifts less 0.2. Fig.~\ref{s2o3dV} shows the distributions of emission lines of the Seyfert 2 galaxies identified by the V\'{e}ron catalog. We find that the emission line luminosities of the Seyfert 2 galaxies identified by the V\'{e}ron catalog still show an extended distributions in high luminous end. This results indicate that the extended tail of the emission line distributions of the Seyfert 2 galaxies is not caused by our selection criteria of Seyfert 2 galaxies.

\begin{figure*}
\begin{center}
\includegraphics[width=0.7\textwidth]{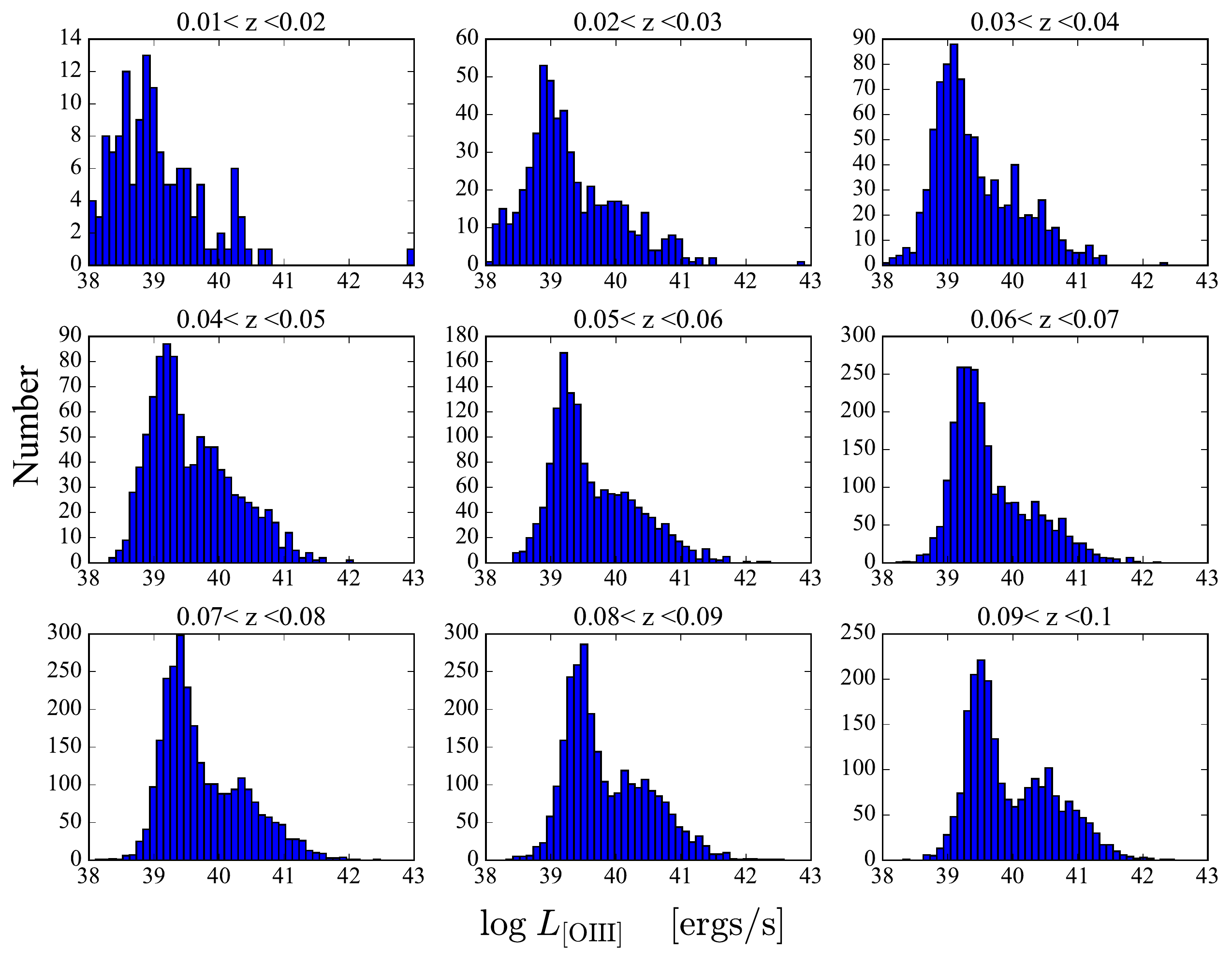}
\caption{$L_{\mathrm{[OIII]}}$ distribution for the sub-sample Seyfert 2 galaxies with S/N$_{\mathrm{[OIII]}}$ and S/N$_{\mathrm{H\alpha}} > 5$ in different redshift ranges.}
\label{sub_s2o3dV}
\end{center}
\end{figure*}

We select a sub-sample of Seyfert 2 galaxies to check whether our results could be affected by some  biases. First, we excluded the sources with FWHM < 100 km~s$^{-1}$, which are not well resolved in the SDSS spectroscopy and might be affected by HII regions. Second, we removed 2337 special sources, which include ROSAT targets and luminous red galaxies. Third, we inspected 55438 spectra of the high luminous Seyfert 2 galaxies by eyes to remove 745 Seyfert 1.9 galaxies. Fourth, we limited the redshifts to be from 0.05 to 0.1 to reduce possible aperture effects. Last, we only considered detection with S/N > 5 for both [OIII] and H$\alpha$ to have sufficient accuracy of measurements. This sub-sample has 8127 sources. We show the L$_{\mathrm{[OIII]}}$ distributions of the sub-sample in Fig.~\ref{sub_s2o3dV} and find that the distributions still have a tail in the high luminosity part. This indicates that our results are not affected by these biases.

\begin{figure}
\begin{center}
\includegraphics[width=0.45\textwidth]{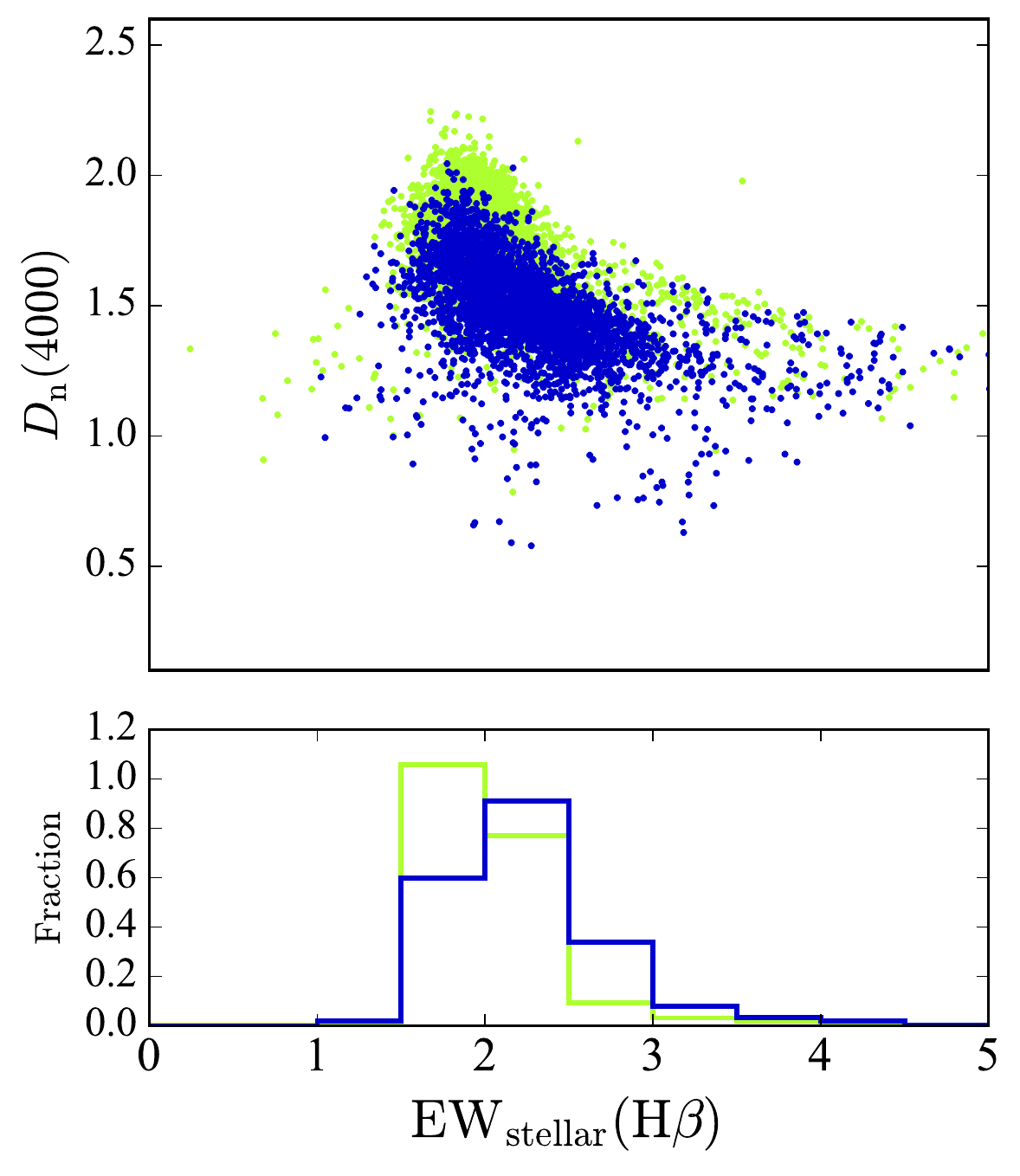}
\caption{H$\beta$ equivalent width from stellar populations versus $D_n(4000)$ {\it{(top)}} and the the fraction of galaxies as a function of H$\beta$ equivalent width {\it{(bottom)}} for the Seyfert 2 galaxies. The stellar equivalent widths were derived by subtracting the AGN H$\beta$ contributions. Green dots represent the sub sample of Seyfert 2 with $\log L_{\mathrm{[OIII]}} \mathrm{[ergs/s]} < 40.125$ and blue dots represent the sub sample of Seyfert 2 with $\log L_{\mathrm{[OIII]}} \mathrm{[ergs/s]} > 40.125$.}
\label{s2ewHb}
\end{center}
\end{figure}
 
 We used the stellar H$\beta$ equivalent width of the Seyfert 2 galaxies as another stellar age tracer \citep{Marcillac06}. The equivalent widths of H$\beta$ are obtained from galSpecLINE table of the SDSS DR10. The EW of the stellar population is derived from the EW of the emission line without accounting for  stellar absorption (REQW) and the EW of the emission line considering stellar absorption (EQW), which was derived from fitting various stellar templates \citep{Tremonti04}. In other words, the stellar EW is equal to REQW subtracted by EQW (REQW-EQW) as mentioned in the table of galSpecLINE \citep{Tremonti04,Brinchmann04}. Fig.~\ref{s2ewHb} shows that both these two Seyfert 2 populations have sources with large and small H$\beta$ equivalent widths; this suggests these two Seyfert 2 populations could have both young and old host galaxies.
 %that the Seyfert 2 galaxies with $\log L_{\mathrm{[OIII]}} \mathrm{[ergs/s]} > 40.125$ have a broader H$\beta$ equivalent width distribution than the Seyfert 2 galaxies $\log L_{\mathrm{[OIII]}} \mathrm{[ergs/s]} < 40.125$ have; this suggests that the high $L_{\mathrm{[OIII]}}$ Seyfert 2 galaxies could have both young and old host galaxies.

\begin{figure*}
\begin{center}
\includegraphics[width=0.9\textwidth]{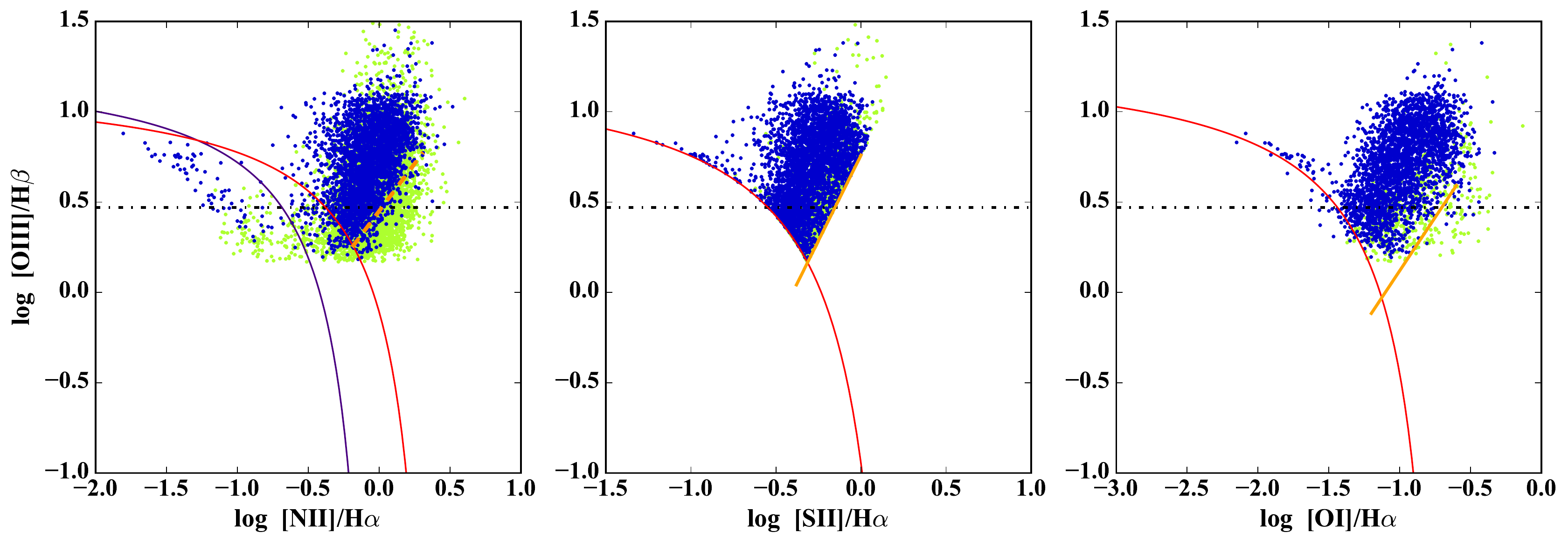}
\caption{Distribution of the sub-sample Seyfert 2 galaxies on the BPT diagram. The red line represents the definition of the starburst limit of \citet{Kewley01}. The orange line represents the Seyfert-LINER line of \citet{Kewley06}. The purple solid line represents the AGN definition of \citet{Kauffmann03}. The black dashed line represents the ratio of [OIII]/$H\beta = 3$. The orange dot line in the left panel represents the Seyfert-LINERS line of \citet{Schawinski07}. The color symbols are the same as in the Fig.~\ref{s2ewHb}. We only show the sources with S/N > {\bf 5} of [NII], [SII], and [OI], respectively. The [NII]/H$\alpha$ diagram has 7855 sources. The [SII]/H$\alpha$ diagram has 5118 sources. The [OI]/H$\alpha$ diagram has 3160 sources.}
\label{s2bpt}
\end{center}
\end{figure*}

We used the BPT diagram \citep{Baldwin81} for the 8127 sources to investigate possible different ionizing mechanisms between our low and high luminosity Seyfert 2 galaxies. The emission lines are constrained with S/N $> 5$ in each diagram of Fig.~\ref{s2bpt}. We found that there are no distinct distributions between the low and high luminosity Seyfert 2 galaxies on these three BPT diagrams. However, we notice that in the diagram of [NII]/H$\alpha$ there are a few Seyfert 2 galaxies located below the line of the definition of an AGN \citep{Kauffmann03}. This suggests that these Seyfert 2 galaxies, which have relatively strong [SII] as shown in the middle plot of Fig.~\ref{s2bpt}, have relatively weak [NII]; these Seyfert 2 galaxies contradicts the early results of \citet{Storchi90}, which suggested that the abundances of [SII] and [NII] should be correlated.

%We noted that the Seyfert 2 sources with $\log L_{\mathrm{H}\alpha} \mathrm{[ergs/s]} > 41.25$ are located in the regions of the BPT diagram similar to that of HBLR Seyfert 2 galaxies in \citet{Yu11}. Besides, the Seyfert 2 galaxies with $\log L_{\mathrm{H}\alpha} \mathrm{[ergs/s]} > 41.25$ also have smaller value of $D_n(4000)$ as the HBLR Seyfert 2 galaxies of \citet{Yu13} do. This suggests that the Seyfert 2 galaxies with $\log L_{\mathrm{H}\alpha} \mathrm{[ergs/s]} > 41.25$ might be related to the HBLR Seyfert 2 galaxies. Although most of the Seyfert 2 with $\log L_{\mathrm{H}\alpha} \mathrm{[ergs/s]} < 41.25$ have $D_n$(4000) more than 1.2 as the non-HBLR Seyfert 2 galaxies of \citet{Yu13} have, the distribution of the Seyfert 2 with $\log L_{\mathrm{H}\alpha} \mathrm{[ergs/s]} < 41.25$ in the BPT diagram is different from that of the non-HBLR Seyfert 2 galaxies; the average value of [OIII]/H$\beta$ of the Seyfert 2 galaxies with $\log L_{\mathrm{H}\alpha} \mathrm{[ergs/s]} < 41.25$ seems to be much weaker than that of the non-HBLR Seyfert 2 galaxies \citep{Yu11}. This indicates that the Seyfert 2 with $\log L_{\mathrm{H}\alpha} \mathrm{[ergs/s]} < 41.25$ might contain weak Seyfert 2 galaxies, LINER, and composite AGNs.

\begin{figure}
\begin{center}
\includegraphics[width=0.4\textwidth]{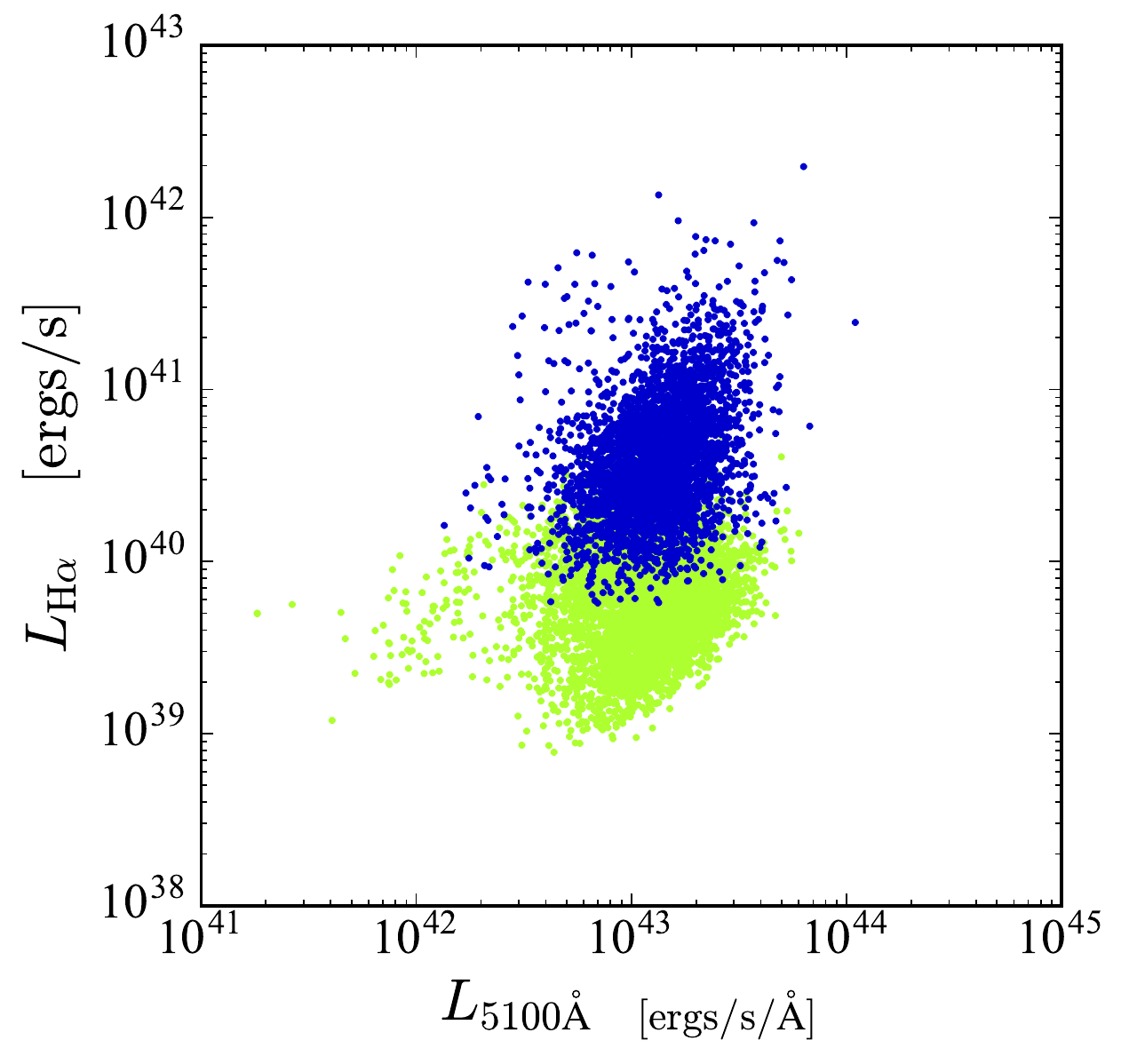}
\caption{Average luminosity from $5050~\AA$ to $5150~\AA$ versus H$\alpha$ luminosity. The color symbols are the same as in the Fig.~\ref{s2ewHb}}
\label{s2cha}
\end{center}
\end{figure}

%\begin{table*}
%\caption{Results of Pearson correlation test for Fig.~\ref{s2cha}}% and Fig.~\ref{s2co3}}
%\label{table: crtsy2}
%\begin{tabular}{lccc}
%\hline
%separation line of $\log L_{\mathrm{[OIII]}}  \mathrm{[ergs/s]} = 40.125$& high luminosity Sy2 & low luminosity Sy2 &  total Sy2\\
%\hline
%Correlation coefficient & -0.029 & 0.047 & -0.0174\\
%P-value & 0.089 & 0.001 & 0.115\\
%\hline
%\end{tabular}
%\end{table*}

We compared the continuum emission of the Seyfert 2 galaxies to check whether the origin of these two distributions of the Seyfert 2 galaxies are related to their continuum emission. We estimated the luminosity density at 5100~\AA, which was estimated by averaging the luminosity from 5050~\AA ~to 5150~\AA ~in the rest-frame. Fig.~\ref{s2cha} presents the results of $L_{5100\AA}$ versus $L_\mathrm{[OIII]}$. We found that although the two populations have different power of $L_{\mathrm{H\alpha}}$, their $L_{5100\AA}$ distributed over similar luminosity range.

\begin{figure}
\includegraphics[width=0.5\textwidth]{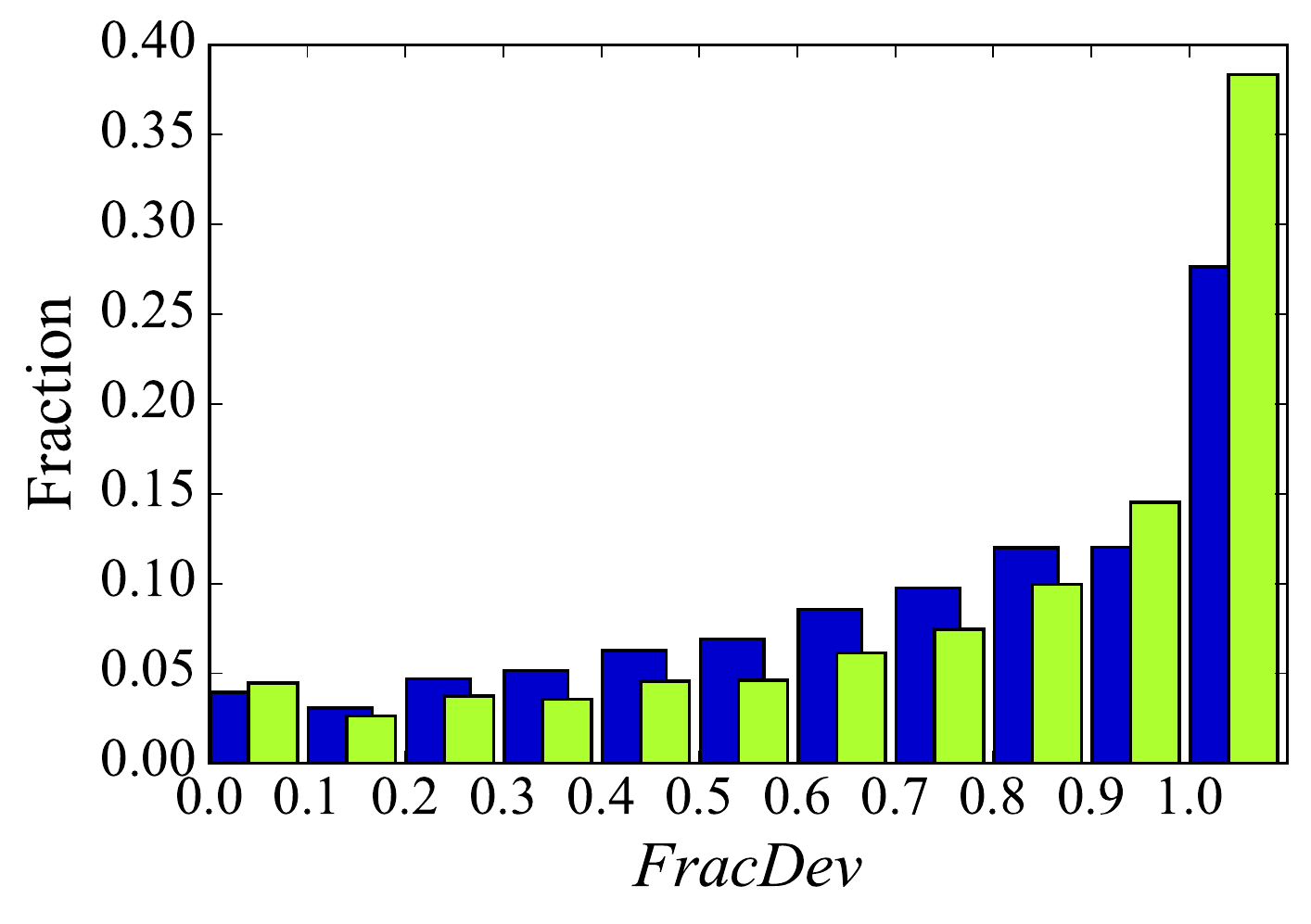}
\caption{{\it FracDev} distributions of the Seyfert 2 galaxies. The color symbols are the same as in the Fig.~\ref{s2ewHb}.}
\label{s2fra}
\end{figure}

%\begin{figure}
%\includegraphics[width=0.5\textwidth]{fracD_sy2.eps}
%\caption{{\it FracDev} distributions of the Seyfert 2 galaxies. The color symbols are the same as in the Fig.~\ref{s2d4k_div}.}
%\label{s2fra_div}
%\end{figure}

We further investigate host galaxy morphology of the sub sample of the Seyfert 2 galaxies. We use {\it FracDev} as a morphology indicator. This parameter describes the bulge contribution in a galaxy and ranges from 0 to 1. A galaxy is more bulge-dominant as {\it FracDev} is closer to 1. Fig.~\ref{s2fra} shows the {\it FracDev} distributions of the Seyfert 2 galaxies. We found that the low and high luminous populations of the Seyfert 2 galaxies have slightly different morphology distributions. We did a Kolmogorov-Smirnov (K-S) test for the {\it FracDev} distributions of the two populations of the Seyfert 2 galaxies. The results of the K-S test have K-S statistic (D) = 0.136 and p-value=1.877$\times10^{-32}$. The extreme low p-value indicates that the probability of the high and low luminous Seyfert 2 galaxies drawn from the same populations is very low. These results suggest that the differences between the high and low luminous Seyfert 2 galaxies are also related to their host galaxy morphology. In other words, the AGN activity of these Seyfert 2 galaxies is related to the morphology of their host galaxies. We also noticed that the {\it FracDev} distribution of the high luminous Seyfert 2 galaxies is similar to the distribution of Seyfert 1 galaxies \citep{Chen17}. This might suggest that the physical properties of the high luminous Seyfert 2 galaxies are similar to those of the Seyfert 1 galaxies.

\section{Summary}
Seyfert 2 galaxies show a correlation between the bulge magnitudes and luminosities of emission lines. We found that there are two groups of Seyfert 2 galaxies with different emission line luminosity distributions. The Seyfert 2 galaxies with larger emission line luminosity have low values of $D_n$(4000), and have larger accretion rates to the central super massive black holes. The significant difference between the two groups of the Seyfert 2 galaxies is related to their host galaxy morphology. 

%We also found that there is a small group of Seyfert 2 galaxies, which show late-type galaxy morphology and have younger stellar populations compared with the other Seyfert 2 galaxies, which are dominated by early-type galaxies and have relatively older stellar populations. 

%We noted that only the Seyfert 2 with $\log L_{\mathrm{[OIII]}} \mathrm{[ergs/s]} > 40.125$ show a galaxy morphology distribution similar to the distribution of Seyfert 1 galaxies.

%We found that the Seyfert 2 galaxies with $\log L_{\mathrm{H}\alpha} \mathrm{[ergs/s]} < 41.25$ show a correlation between their [OIII] and continuum emission whereas there is no such a relation for the Seyfert 2 galaxies with $\log L_{\mathrm{H}\alpha} \mathrm{[ergs/s]} > 41.25$. 

\section*{Acknowledgements}

This work is supported by the Ministry of Science and Technology of Taiwan (grant MOST 107-2119-M-008-009-MY3). We thank L. Kewley, P.~C.~Yu, Z.~Y.~Chen, J.~C.~Huang, R. Ruffini, and P. Giommi for discussion and comments.

Funding for SDSS-III has been provided by the Alfred P. Sloan Foundation, the Participating Institutions, the National Science Foundation, and the U.S. Department of Energy Office of Science. The SDSS-III web site is http://www.sdss3.org/. SDSS-III is managed by the Astrophysical Research Consortium for the Participating Institutions of the SDSS-III Collaboration including the University of Arizona, the Brazilian Participation Group, Brookhaven National Laboratory, Carnegie Mellon University, University of Florida, the French Participation Group, the German Participation Group, Harvard University, the Instituto de Astrofisica de Canarias, the Michigan State/Notre Dame/JINA Participation Group, Johns Hopkins University, Lawrence Berkeley National Laboratory, Max Planck Institute for Astrophysics, Max Planck Institute for Extraterrestrial Physics, New Mexico State University, New York University, Ohio State University, Pennsylvania State University, University of Portsmouth, Princeton University, the Spanish Participation Group, University of Tokyo, University of Utah, Vanderbilt University, University of Virginia, University of Washington, and Yale University.

%%%%%%%%%%%%%%%%%%%%%%%%%%%%%%%%%%%%%%%%%%%%%%%%%%

%%%%%%%%%%%%%%%%%%%% REFERENCES %%%%%%%%%%%%%%%%%%

% The best way to enter references is to use BibTeX:

%\bibliographystyle{mnras}
%\bibliography{example} % if your bibtex file is called example.bib

% Alternatively you could enter them by hand, like this:
% This method is tedious and prone to error if you have lots of references

%%%%%%%%%%%%%%%%%%%%%%%%%%%%%%%%%%%%%%%%%%%%%%%%%%

%%%%%%%%%%%%%%%%%%%%%%%%%%%%%%%%%%%%%%%%%%%%%%%%%%

% Don't change these lines
\bsp	% typesetting comment
\label{lastpage}
\end{document}